# Experimental study on an S-band near-field microwave magnetron power transmission system on hundred-watt level

Biao Zhang, Wan Jiang, Yang Yang, Chengyang Yu, Kama Huang and Changjun Liu ⓘ*

*School of Electronics and Information Engineering, Sichuan University, Chengdu 610064, China*

A multi-magnetron microwave source, a metamaterial transmitting antenna, and a large power rectenna array are presented to build a near-field 2.45 GHz microwave power transmission system. The square 1 m² rectenna array consists of sixteen rectennas with 2048 Schottky diodes for large power microwave rectifying. It receives microwave power and converts them into DC power. The design, structure, and measured performance of a unit rectenna as well as the entail rectenna array are presented in detail. The multi-magnetron microwave power source switches between half and full output power levels, i.e. the half-wave and full-wave modes. The transmission antenna is formed by a double-layer metallic hole array, which is applied to combine the output power of each magnetron. The rectenna array DC output power reaches 67.3 W on a 1.2 Ω DC load at a distance of 5.5 m from the transmission antenna. DC output power is affected by the distance, DC load, and the mode of microwave power source. It shows that conventional low power Schottky diodes can be applied to a microwave power transmission system with simple magnetrons to realise large power microwave rectifying.

**Keywords:** microwave rectifying; microwave power transmission; rectenna; near-field

## 1. Introduction

Microwave Power Transmission (MPT) (Brown, 1984) is a key technology for wireless power transmission (WPT) and has been widely studied due to its advantages such as low loss, high power, and all-weather working, etc. An MPT system conveys the electrical power wirelessly by microwave or laser. The microwave frequency band is located at the atmosphere radio window which allows the electromagnetic radiation to transmit with low dissipation. The output power of microwave has been rapidly improved with the technique development of magnetrons and high-power solid-state devices, which makes an MPT system closer to practical applications.

Recently, a few types of rectennas with linear and circular polarisation in various shapes were reported (Olgun, Chen, & Volakis, 2011; Strassner & Chang, 2002; Strassner & Chang, 2003). Most of them focused on either a single rectenna or a low power rectenna array at a fixed transmission distance. In an MPT system, a rectenna array at least with watt level is convenient and useful (He & Liu, 2009). The power capacitance is mainly determined by the diodes applied in the rectenna. Schottky diodes are a type of commonly used rectifying diodes. However, low power capacitance of a Schottky diode limits its applications to large power conditions. Apart from power capacitance, transmission distance is another essential factor which influences the system transmission efficiency and output DC power.

---

*Corresponding author. Email: cjliu@ieee.org

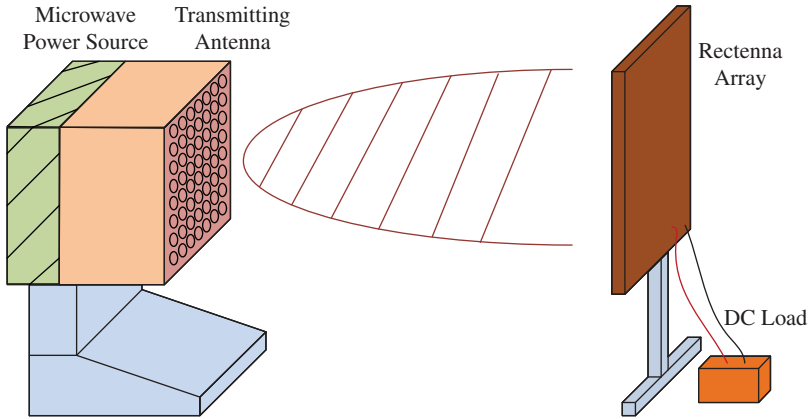

Figure 1.   Configuration of the MPT system.

As shown in Figure 1, an MPT system comprising a microwave power source, a transmitting antenna, a rectenna array, and a DC load (Shinohara & Matsumoto, 1998) realises transmitting microwave power from one location to another wirelessly. The transmitted power is received and converted to DC power by the rectenna array. In the receiving system, rectenna is a key component which determines both the power capability and the MW-to-DC conversion efficiency of entire system. A rectenna is generally composed of an antenna to receive microwave power and a rectifier to convert the microwave power into DC power. The proposed design in this paper shows a single rectenna is able to export 11.4 W DC power with 55% MW-to-DC conversion efficiency.

In order to enhance the power capacitance of an MPT system, some improvements are adopted. A multi-magnetron microwave power source and a metamaterial transmitting antenna are applied to microwave power transmitting. A large power square rectenna array is built at 2.45 GHz for microwave power transmission system. It is a 1 m$^2$ array consisting of sixteen rectennas with 2048 Schottky diodes. The multi-magnetron microwave power source can switch between half and full output power levels. The rectenna array's DC output power reaches 67.3 W on a 1.2 Ω DC load at a distance of 5.5 m from the transmitting antenna. The DC output power is affected by distance, DC load, and the mode of microwave power source. It shows that conventional low power Schottky diodes can be applied to a microwave power transmission system with simple magnetrons to realise large power microwave rectifying.

## 2.   Microwave source and transmitting antenna

The microwave source has four magnetrons and four horn antennas. The output microwave power of four magnetrons is combined by a metamaterial lens in front of the horn antennas. The metamaterial lens is formed by a double-layer metallic holes array for high power microwave applications (Liu, Yang, & Huang, 2013). By increasing the effective electron mass of metamaterial, the effective plasma frequency reduces to microwave band (Pendry, Holden, & Robbins, 1998) while effective dielectric constant of the material depends both on working frequency and effective plasma frequency. When operating frequency approaches the effective plasma frequency, effective dielectric constant is close to zero (Pendry et al., 1998) and the double-layered metallic holes array becomes equivalent to a zero-refractive-index material (Figure 2).

In the experiments, four Panasonic 2M244-M1 magnetrons which are respectively powered by four independent high voltage source and four 15 dBi horn antennas are employed as microwave sources and transmitting antennas. Frequency of the high-voltage AC source is 50 Hz and the magnetron usually only works in the positive half-period. Figure 3 shows a typical frequency spectrum of a 2.45 GHz oven magnetron (Mitani, Kawasaki, Shinohara, & Matsumoto, 2009). Under normal operating conditions, the magnetron does not obtain a high quality spectrum, which means the frequency is not stable.

By testing the antenna pattern, the half-power beamwidth decreases from 28º to 15º and the gain increases by 1.7 dB at 2.47 GHz compared to non-metamaterials (Liu et al., 2013). High gain and small half-power beamwidth enhance the transmission efficiency and reduce the area of receiving rectenna array.

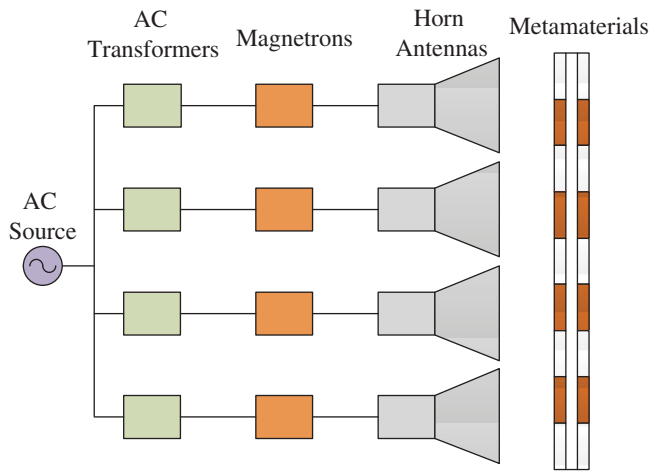

Figure 2.   Schematic of the transmitting antenna.

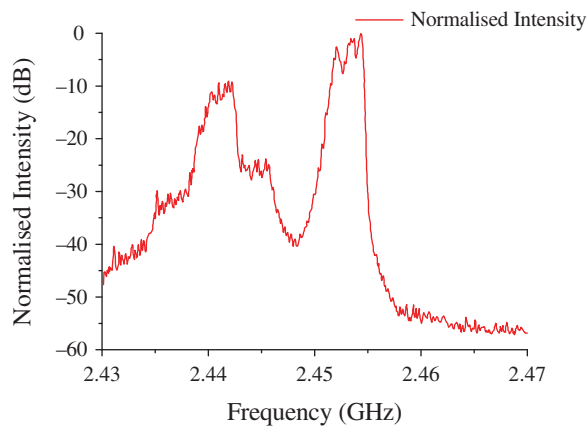

Figure 3.   Typical frequency spectrum of a 2.45 GHz oven magnetron.

## 3. Rectenna of the MPT system

A new concentrated rectenna is developed for the near-field MPT system. It is composed of a microstrip grid array antenna and a concentrated rectifier composed of 32 HSMS-282P diode bridges for power capacity enhancement, which are connected by an inner conductor of a 50 Ω coaxial cable. Figure 4 shows a see-through view of concentrated rectenna which has a multi-layer structure consisting of an antenna layer, an air layer, a ground layer and a circuit layer. Two F4B-2 dielectric substrates with a dielectric constant of 2.65 and thickness of 1 mm are inserted between the antenna layer and air layer, and between the ground and circuit layer. An inner conductor of 50 Ω coaxial cable passes through middle layers to connect the top antenna layer and the bottom circuit layer.

### 3.1. *Receiving antenna*

The microstrip grid array antenna (Kraus & Marhefka, 2002) is composed of 7 rectangle patches and each patch is linked by microstrips to match the impedance and transfer the power. It is fed simply at one point by a 50 Ω coaxial cable with inner conductor through a hole in the ground plane, and with the outer conductor bonded to the ground plane. Figure 5 presents the photograph of the grid array antenna while Figure 6 shows the measured |S$_{11}$| and pattern of the antenna. Return loss of the antenna is 20 dB at the frequency of 2.45 GHz and the half-power beamwidth (HPBW) is equal to 25°.

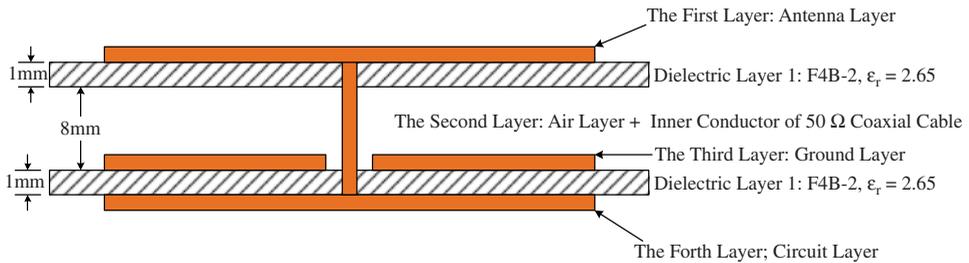

Figure 4.   See-through illustration of the rectenna for multi-layer structure.

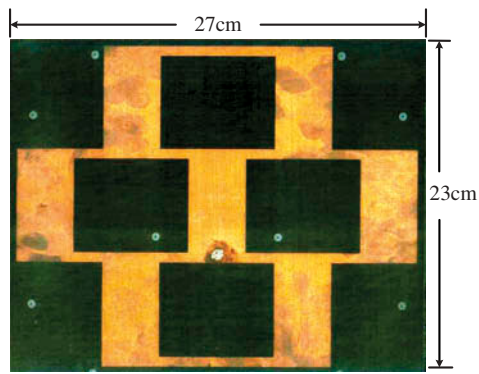

Figure 5.   The photograph of the grid array antenna.

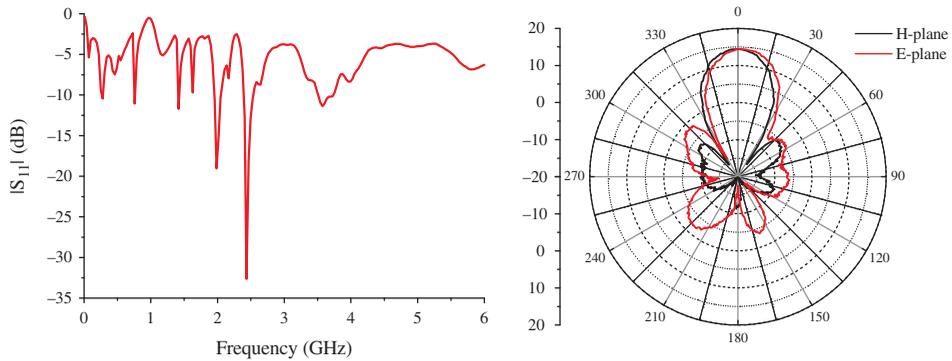

Figure 6. Measured return loss and pattern of the receiving antenna.

### 3.2. *Rectifying circuit*

Figure 9(a) illustrates the photograph of the concentrated rectifying circuit which includes eight branches linked to central feed point with a 9 dB power divider. Each branch is based on four HSMS-282P Schottky diode bridges to achieve rectification, where each piece of HSMS-282P diode bridge packages 4 HSMS-282 internal Schottky diodes. For the 9 dB power divider, the transmission coefficients between the input and output ports are near −9 dB and insertion loss is less than 0.3 dB, while the reflection coefficient of input port is less than −25 dB and the isolation between the outputs ports is greater than 18 dB at 2.45 GHz. For each branch, as shown in Figure 7, the highest conversion efficiency is 69% at 30 dBm microwave input power (Zhang, Zhao, Yu, Huang, & Liu, 2011). In order to avoid overheating of diodes of each branch, two half-wavelength microstrip lines placed between the diodes are introduced in the design to separate them. The half-wavelength microstrip line duplicates the impedance of one port to another (Pozar, 2005).

$$Z_{in} = Z_0 \frac{Z_L + jZ_0 \tan \beta l}{Z_0 + jZ_L \tan \beta l} \tag{1}$$

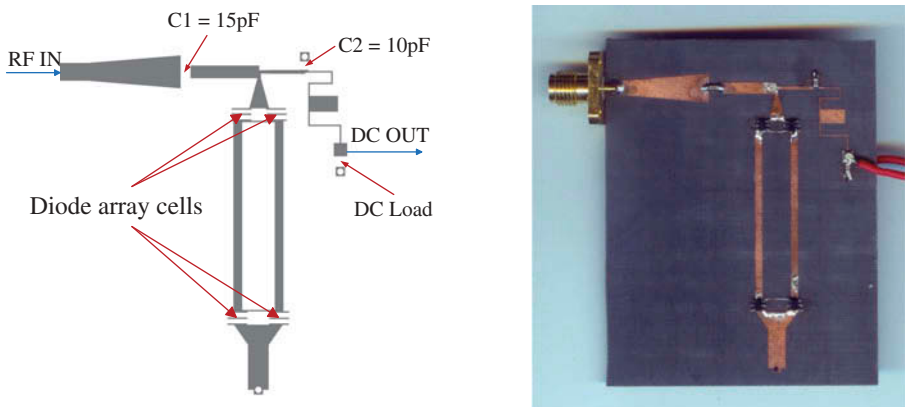

Figure 7. The layout and the photograph of the branch circuit.

where $Z_{in}$ is the input impedance, $Z_0$ is the transmission line characteristic impedance, $Z_L$ is the load impedance, and $\beta$ is the propagation constant.

If $l = \lambda/2$, Equation (1) shows that

$$Z_{in} = Z_L. \tag{2}$$

As shown in Figures 7 and 8, sixteen diodes have been applied to forming a 4 × 4 array. Adding two half-wavelength microstrip lines not only helps the heat dissipation, but also compensates the parasite effects of diodes without changing the impedance of a single diode.

Temperature of diodes rises when they work in the rectifying circuit. In order to detect whether the diodes work in a stable state, temperature of the circuit was monitored at the input power of 40 dBm. Figure 9(b) shows the temperature distribution of the circuit after 30 minutes working, where the white parts in the figure represent high temperature. Compared to the rectifying circuit, heat sources are the diodes with an approximately uniform distribution of temperature, meaning the diodes work in a similar condition with the temperature of 32.8°C. At this temperature, working condition of the diodes is stable.

The rectifying efficiency with respect to input power at different load is shown in Figure 10. The conversion efficiency is higher than 65% at 76.5 Ω and 81.6 Ω when the input power is between 39 and 41 dBm. With the increasing in input power, conversion efficiency increases simultaneously. The highest efficiency is achieved 68.1% at 76.5 Ω load of 41 dBm input power. Power capacitance of this large power rectifying circuit is 43 dBm while the rectifying efficiency is 55%.

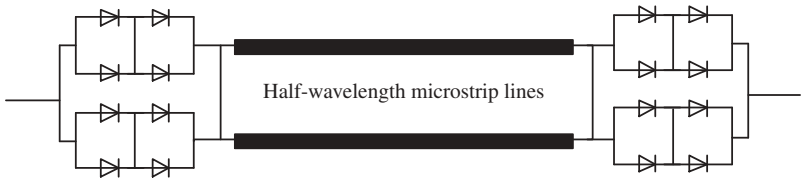

Figure 8. Structure of a 4 × 4 diode array.

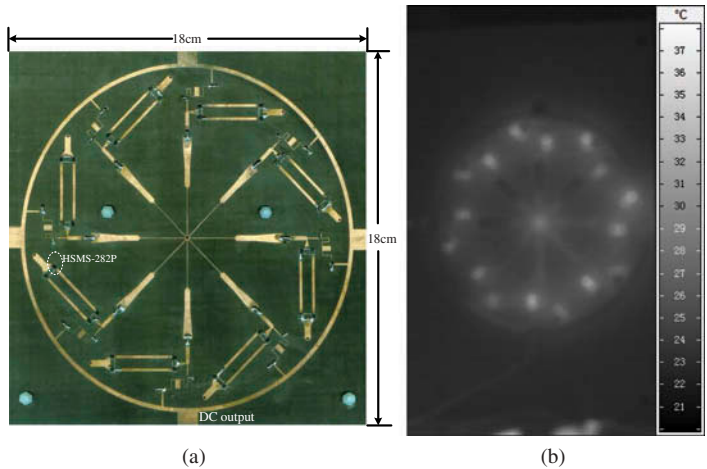

(a)        (b)

Figure 9. The photograph and the temperature distribution of the rectifying circuit.

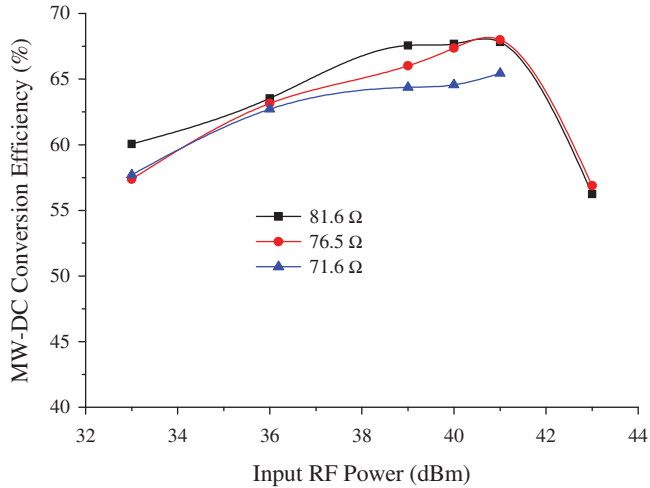

Figure 10.   The measured rectifying efficiency with respect to the input power at different load.

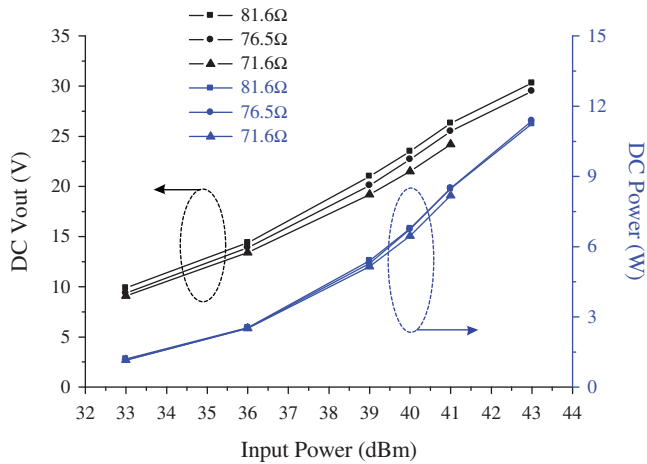

Figure 11.   The measured DC voltage and power respect to the input MW power.

Figure 11 illustrates the DC voltage and power with respect to input MW power. Six lines indicate different levels of the DC load. DC voltage and power increase linearly for all three DC load levels. The highest voltage is 30 V at 81.6 Ω and the maximum DC power is 11.4 W at 76.5 Ω.

## 4.   Experimental results of the rectenna array

The field around an antenna is divided into two principal regions: one is near the antenna called the near-field, and the other is at a large distance called the far-field. The boundary between them is at a ratio

$$R = \frac{2L^2}{\lambda} \tag{3}$$

where $L$ is the maximum dimension of the antenna.

In the experiment, radius of the near far-field boundary is equal to 33 m from Equation (3) when $L = \sqrt{2}$m. The maximum distance between the transmitting antenna and receiving antenna is 8 m. Thus, the receiving antenna is not in the far-field of transmitting antenna. Therefore, the microwave power transmission system works in the near-field.

### 4.1. *Rectenna array used in the experiment*

In the near-field MPT experiment, a sub-array system for the rectenna array is employed. Each sub-array consists of two concentrated rectennas electrically connected in parallel. The rectenna array with the size of 1.1 m × 0.9 m consists of 2 × 4 = 8 sub-arrays, namely, 16 rectenna elements, as shown in Figure 12. Such sub-array systems provide flexibility with regard to mechanical structure and electrical connection. In order to compare, we tested not only the single sub-array but also the entire rectenna array in this experiment.

### 4.2. *Experiment results of single sub-array*

The transmitting antenna contains four independent groups of magnetron and horn antenna. Because the frequency of the high voltage supplied to the magnetron is 50 Hz, the magnetron only works in the positive half period. In order to make the transmitting antenna generate power entire period, the positions of hot and neutral lines are exchanged at the input of half magnetron's high-voltage transformer. Thus, for entire period of 50 Hz, two magnetrons work together.

At first, a single sub-array rectenna with a distance of 8 m between the source and rectenna element was measured. The two magnetrons, as the microwave power source, worked in three modes. In Mode I, single magnetron worked as power source and the output of DC voltage was tested. In Mode II, two magnetrons which worked in the same half period of 50 Hz were applied. In Mode III, one worked in half period while the other in the other half. The waveforms are shown in Figure 13(a–c), measured by Tektronix DPO 7254 Digital Phosphor Oscilloscope and ATTEN ATZ3710A DC electronic load.

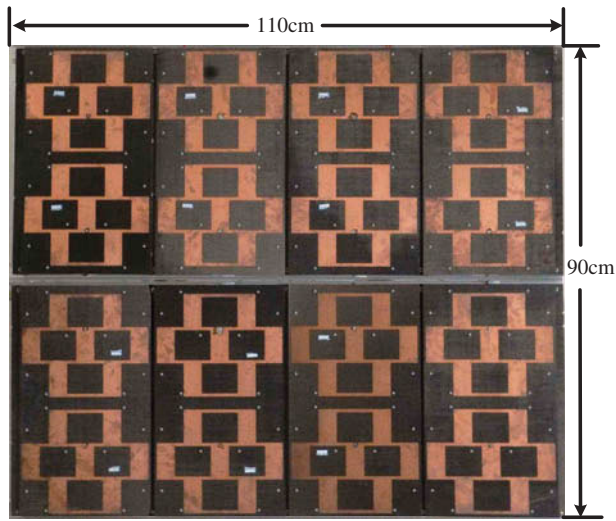

Figure 12. Rectenna array used in the experiment.

All the output waveforms shown in Figure 13 approximate to square wave signal, even if the two magnetrons work in opposite half cycles. In Figure 13(a), the amplitude of output voltage is 6.96 V, meanwhile the root mean square (RMS) of the voltage is 3.84 V. In Figure 13(b), the amplitude equals 9.04 V and the RMS equals 5.15 V. In Figure 13(c), the amplitude and the RMS are 7.44 V and 5.74 V, respectively.

In Mode I and III, the amplitudes are close, but half the difference of the squared RMSs. In Mode II and III, the RMSs are close, but half the difference of the squared amplitudes. These results show that the two magnetrons working in opposite half cycles could reduce the peak voltage but the RMS remains unchanged even though it is enhanced. Moreover, reducing the peak voltage is an effective way to avoid the diode damage. In Figure 13(c), due to oscillating delay of resonator of magnetron, space is generated between the waveforms. In the event where the magnetron could work on full-wave mode, the RMS would be higher and the waveform of the voltage more stable.

### 4.3. *Experiment results of rectenna array*

The conclusion safely drawn from the above experiment is that the two magnetrons work better in two opposite half periods than in the same half period. Therefore, the entire rectenna array was tested with four magnetrons with two working in half period and the remaining two in the other half. At the output terminal of the array, a voltage regulated circuit was added to safeguard the diodes in the rectifying circuit and make the ripple of DC current smoother.

When the circuit is added to the entire array at 2 Ω DC load, the output DC voltage waveform is shown in Figure 14. At this time, the transmitting antenna works at Mode III, which is mentioned in single sub-array experiment. The RMS of output DC voltage is equal to 8.12 V.

Figures 15 and 16 show the DC voltage and power in respect to the load at different distances. When the load increases, the output DC voltage and power rise as well. Once the resistance reaches a certain value, voltage stabilises at 12 V while the power rapidly declines due to self-protection circuit of rectifier. The maximum DC power archives 67.3 W with 1.2 Ω DC load at a distance of 5.5 m.

### 5. Conclusion

In this near-field microwave power transmission experiment, a transmitting antenna composed of four groups of magnetron and horn antenna with metamaterials is built. Then a 1.1 m × 0.9 m rectenna array with each element connected in parallel is constructed. It is composed by 16 sub-arrays and every sub-array contained a microstrip grid array antenna and a concentrated rectifying circuit.

In order to examine the output characteristics of transmitting antenna, a single sub-array rectenna is tested with two magnetrons on. The MPT system is measured at different distances and results show that the maximum DC power archives 67.3 W with 1.2 Ω DC load at the distance of 5.5 m.

The following conclusions are drawn from the experiment results:

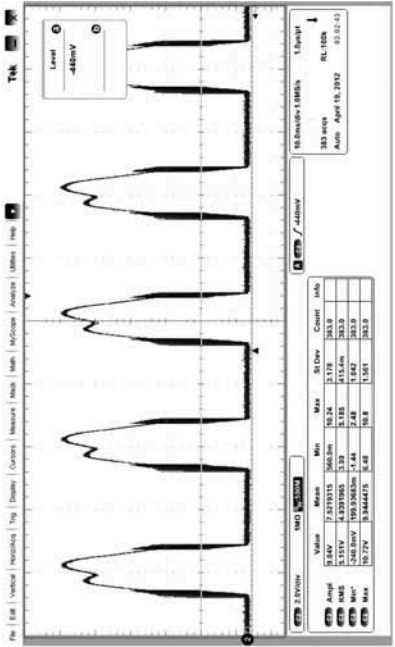

(a)

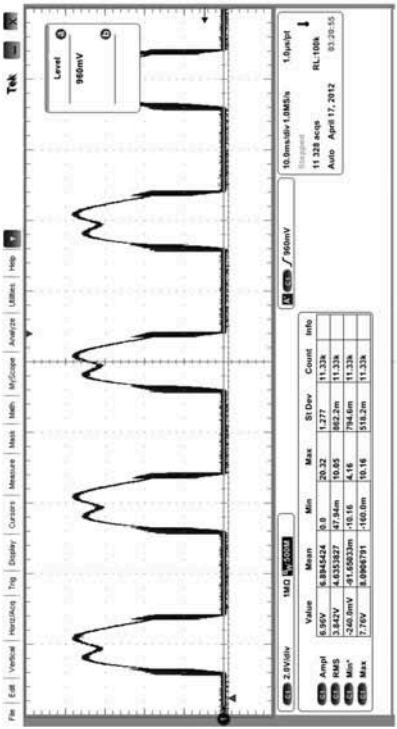

(b)

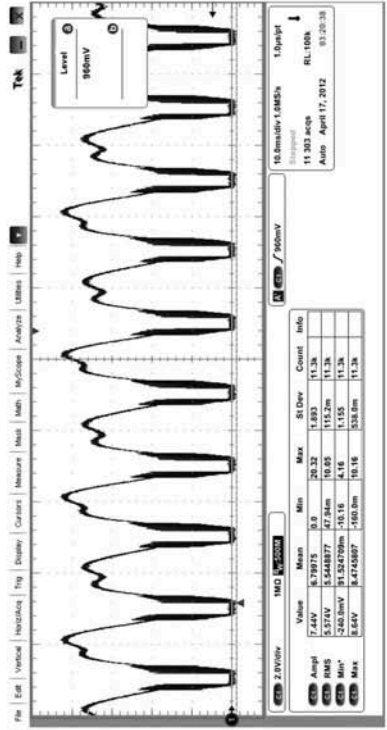

(c)

Figure 13.    The output voltage waveform of the single sub-array rectenna.

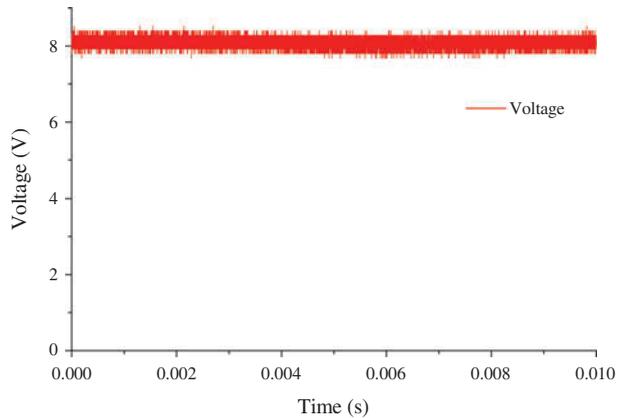

Figure 14.   The output voltage waveform of the entire rectenna array with the voltage regulated circuit.

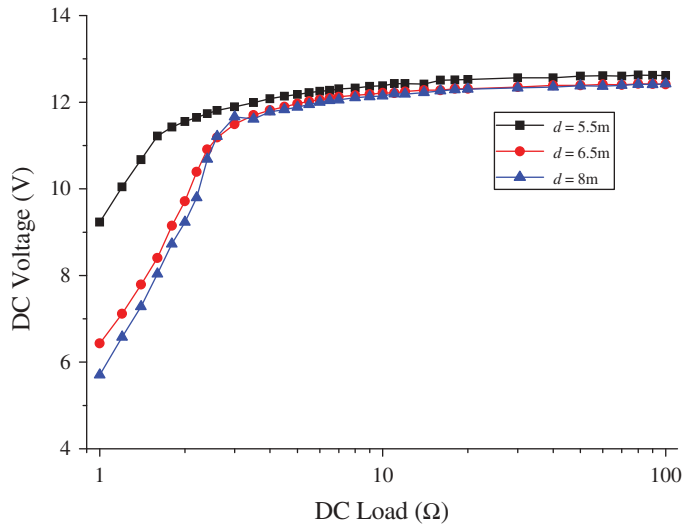

Figure 15.   The DC voltage of the rectenna array as a function of DC load.

(1) The low power Schottky diode can be used in large power rectifying circuit and MPT system.

(2) The requirement of signal quality is lower in microwave power transmission than in the communication system.

(3) In terms of transmission efficiency, a full-wave microwave source is better than a half-wave one.

The significant directions of the future work would be: (1) researching multiple frequency-phase locked magnetrons in full-wave mode for microwave power source; (2) improvement of MW to DC efficiency and power capacitance of single rectifying circuit; (3) designing high radiation efficiency antenna or antenna array for transmitting and receiving antenna; and (4) characteristics of the series and/or parallel connected rectenna unit arrays (Sakamoto, Ushijima, Nishiyama, Aikawa, & Toyoda, 2013).

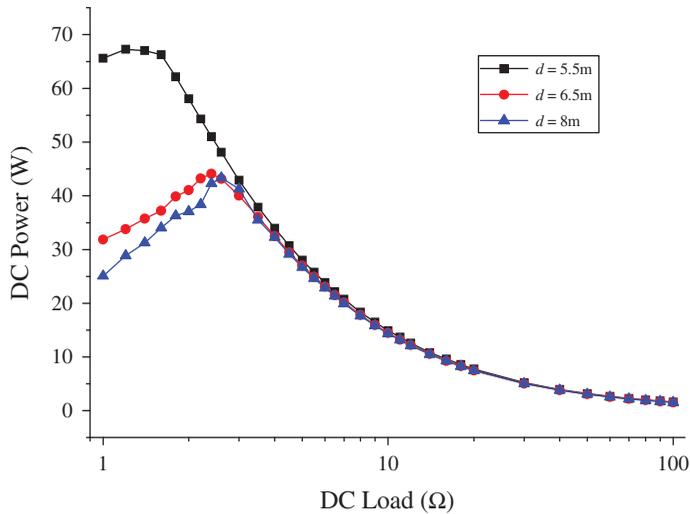

Figure 16. The DC power of the rectenna array as a function of DC load.

The low energy and far-field microwave power transmission has been well studied in last several decades (Pinuela, Mitcheson, & Lucyszyn, 2013; Visser, 2013). However, there were little studies on near-field and large energy transmission (Huang, Zhang, Chen, Huang, & Liu, 2013). Apart from half-wave and full-wave rectifier (Imai, Tamaru, Fujimori, Sanagi, & Nogi, 2011), the full-wave microwave and half-wave source need to be studied for large power MPT system. The work in this paper will be an exploration of transmitting energy without wire but by microwave. In future, we would continue to study the MPT system and try to use it to serve the communication system (Yang, Jiang, Elsherbeni, Yang, & Wang, 2013).

## Funding

This work was supported in part by the NSFC [61271074], 973 program [2013CB328902], [NCET-12-0383].

## ORCID

*Changjun Liu*  http://orcid.org/0000-0003-0079-6793

## References

Brown, W. C. (1984). The history of power transmission by radio waves. *IEEE Transactions on Microwave Theory and Techniques, 32*(9), 1230–1242. doi:10.1109/TMTT.1984.1132833

He, Q., & Liu, C. (2009). An enhanced microwave rectifying circuit using HSMS-282. *Microwave and Optical Technology Letters, 51*(4), 1151–1153. doi:10.1002/mop.24237

Huang, W., Zhang, B., Chen, X., Huang, K.-M., & Liu, C. (2013). Study on an S-band rectenna array for wireless microwave power transmission. *Progress in Electromagnetics Research, 135*, 747–758. doi:10.2528/PIER12120314

Imai, S., Tamaru, S., Fujimori, K., Sanagi, M., & Nogi, S. (2011, May 12–13). *Efficiency and harmonics generation in microwave to DC conversion circuits of half-wave and full-wave rectifier types.* IMWS-IWPT2011, Kyoto. doi:10.1109/IMWS.2011.5877081

Kraus, J. D., & Marhefka, R. J. (2002). *Antennas for all applications* (3rd ed.). Upper Saddle River, NJ: McGraw Hill.

Liu, Q., Yang, Y., & Huang, K. M. (2013). Design and applications of metamaterials with high power microwave. *Journal of Terahertz Science and Electronic Information Technology*, *11*(2), 245–249.

Mitani, T., Kawasaki, H., Shinohara, N., & Matsumoto, H. (2009, April). A study of oven magnetrons toward a transmitter for space applications. In *Vacuum electronics conference, 2009. IVEC'09. IEEE International* (pp. 323–324). Rome: IEEE. doi:10.1109/IVELEC.2009.5193408

Olgun, U., Chen,-C.-C., & Volakis, J. L. (2011). Investigation of rectenna array configurations for enhanced RF power harvesting. *IEEE Antennas and Wireless Propagation Letters*, *10*, 262–265. doi:10.1109/LAWP.2011.2136371

Pendry, J. B., Holden, A. J., & Robbins, D. J. (1998). Low frequency plasmons in thin wire structures. *Journal of Physics: Condensed Matter*, *10*(22), 4785–4809.

Pinuela, M., Mitcheson, P. D., & Lucyszyn, S. (2013). Ambient RF energy harvesting in urban and semi-urban environments. *IEEE Transactions on Microwave Theory and Techniques*, *61*(7), 2715–2726. doi:10.1109/TMTT.2013.2262687

Pozar, D. M. (2005). *Microwave engineering* (3rd ed., p. 77). New York, NY: Wiley.

Sakamoto, T., Ushijima, Y., Nishiyama, E., Aikawa, M., & Toyoda, I. (2013). 5.8 GHz series/parallel connected rectenna array using expandable differential rectenna units. *IEEE Transactions on Antennas and Propagation*, *61*(9), 4872–4875. doi:10.1109/TAP.2013.2266316

Shinohara, N., & Matsumoto, H. (1998). Experimental study of large rectenna array for microwave energy transmission. *IEEE Transactions on Microwave Theory and Techniques*, *46*(3), 261–268. doi:10.1109/22.661713

Strassner, B., & Chang, K. (2002). 5.8 GHz circularly polarized rectifying antenna for wireless microwave power transmission. *IEEE Transactions on Microwave Theory and Techniques*, *50*(8), 1870–1876. doi:10.1109/TMTT.2002.801312

Strassner, B., & Chang, K. (2003). Highly efficient C-band circularly polarized rectifying antenna array for wireless microwave power transmission. *IEEE Transactions on Antennas and Propagation*, *51*(6), 1347–1356. doi:10.1109/TAP.2003.812252

Visser, H. J. (2013, January). Far-field RF energy transport. In *Radio and Wireless Symposium (RWS), 2013 IEEE* (pp. 34–36). Austin, TX: IEEE. doi:10.1109/RWS.2013.6486632

Yang, -X.-X., Jiang, C., Elsherbeni, A. Z., Yang, F., & Wang, Y.-Q. (2013). A novel compact printed rectenna for data communication systems. *IEEE Transactions on Antennas and Propagation*, *61*(5), 2532–2539. doi:10.1109/TAP.2013.2244550

Zhang, B., Zhao, X., Yu, C. Y., Huang, K. M., & Liu, C. (2011). A power enhanced high efficiency 2.45GHz rectifier based on diode array. *Journal of Electromagnetic Waves and Applications*, *25*, 765–774. doi:10.1163/156939311794827159